\documentclass[pra,twocolumn,superscriptaddress,amsmath,amssymb,floatfix]{revtex4-2} 
\usepackage{graphicx} 
\usepackage{bm}       
\usepackage[colorlinks=true,linkcolor=blue,urlcolor=blue,citecolor=blue,linkcolor=blue,pdfstartview=FitH]{hyperref}

\usepackage{enumerate}
\usepackage{amsmath}
\usepackage{xcolor}
\usepackage{cases}
\usepackage{microtype}
\usepackage{siunitx}
\usepackage{subdepth}
\usepackage[utf8]{inputenc}
\usepackage[T1]{fontenc}
\usepackage{txfonts} 
\newcommand{\diff}{\mathrm{d}}
\newcommand{\Nv}{N_\mathrm{v}}

\begin{document}

\title{Fluctuation theorem anomaly in a point-vortex fluid}

\author{Rama Sharma}
\affiliation{Optical Sciences Centre, Swinburne University of Technology, Melbourne 3122, Australia}
\author{Tapio P. Simula}
\affiliation{Optical Sciences Centre, Swinburne University of Technology, Melbourne 3122, Australia}
\author{Andrew J. Groszek}
\affiliation{ARC Centre of Excellence for Engineered Quantum Systems, School of Mathematics and Physics, University of Queensland, St. Lucia, QLD 4072, Australia}
\affiliation{ARC Centre of Excellence in Future Low-Energy Electronics Technologies, School of Mathematics and Physics, University of Queensland, Saint Lucia QLD 4072, Australia}

\begin{abstract}
The second law of thermodynamics posits that in closed macroscopic systems the rate of entropy production must be positive. However, small systems can exhibit negative entropy production over short timescales, seemingly in contradiction with this law. The fluctuation theorem quantitatively connects these two limits, predicting that entropy producing trajectories become exponentially dominant as the system size and measurement time are increased. Here we explore the predictions of the fluctuation theorem for a fluid of point-vortices, where the long-range interactions and existence of negative absolute temperature states provide an intriguing test bed for the theorem. Our results suggest that while the theorem broadly holds even at negative absolute temperatures, the long-range interactions inherent to the vortex matter lead to anomalously large entropy production over short time intervals. The predictions of the fluctuation theorem are only fully recovered when sufficient noise is introduced to the dynamics to overwhelm the vortex--vortex interactions.
\end{abstract}

\maketitle

\section{Introduction}

The second law of thermodynamics states that in a closed macroscopic system entropy will only ever increase~\cite{boltzmann1974second, adkins_equilibrium_1983, attard_non-equilibrium_2012}. However, as identified by Loschmidt~\cite{loschmidt1876zustand}, the question arises how such irreversible behaviour can emerge from the microscopic equations of motion, which are themselves reversible. In mathematical terms, why is it that a system's phase space trajectory has a preferred direction towards a higher entropy state, when the corresponding time-reversed trajectory would appear to be equally likely to occur? Fluctuation theorems (FTs) provide one approach for resolving this paradox. They predict that a physical system can always exhibit both entropy-producing and entropy-reducing trajectories through phase space, but that the latter are exponentially suppressed as the system size or evolution time is increased. These theorems therefore bridge the gap between the reversible microscopic and the irreversible macroscopic dynamics, and recover the familiar second law behaviour in the thermodynamic limit.

FTs can be categorised into two broad classes: Evans--Searles type FTs, which predict the statistics of the entropy produced in systems perturbed from equilibrium by a constant driving~\cite{evans_probability_1993, evans_equilibrium_1994, gallavotti_dynamical_1995, gallavotti_dynamical_1995-1,evans_fluctuation_2002, sevick_fluctuation_2008}, and Crooks type FTs, which predict the statistics of work fluctuations in systems with time-dependent driving~\cite{jarzynski_nonequilibrium_1997, crooks_entropy_1999}. Importantly, these theorems are not restricted to equilibrium systems, providing one of the few analytical thermodynamic predictions valid in the nonequilibrium regime. A wealth of experimental verification now exists for both forms of FT~\cite{wang_experimental_2002, liphardt_equilibrium_2002, collin_verification_2005, garnier_nonequilibrium_2005, schuler_experimental_2005, douarche_work_2006, tietz_measurement_2006, utsumi_bidirectional_2010}, cementing their importance in our understanding of statistical physics.

In this work we consider the statistical properties of a fluid of point-vortices. This system consists of a collection of point-like vortex `particles', which interact with each other via long-range Coulombic velocity fields. Point-vortices were first considered as a toy model for two-dimensional (2D) fluids, but have more recently been found to accurately describe the dynamics of quantised vortices in 2D superfluid Bose--Einstein condensates~\cite{navarro_dynamics_2013, billam_onsager-kraichnan_2014, simula_emergence_2014, groszek_vortex_2018, gauthier_giant_2019, johnstone_evolution_2019, groszek_crossover_2021, reeves_turbulent_2022}. One remarkable feature of this system is its ability to exhibit negative absolute temperature states in any bounded container, as identified by Onsager~\cite{onsager1949statistical}. In this negative temperature regime, the vortices tend to form same-sign clusters~\cite{reeves_inverse_2013, simula_emergence_2014, billam_onsager-kraichnan_2014, groszek_onsager_2016}, maximising the kinetic energy of the flow field while reducing the configurational entropy. These negative temperature states have now been realised experimentally in ultracold gases~\cite{gauthier_giant_2019, johnstone_evolution_2019}, and more recently in exciton--polariton condensates~\cite{panico_onset_2023}.

Here we investigate the applicability of the Evans--Searles FT to the point-vortex fluid, given its unusual features of long-range interacting particles and negative absolute temperatures. In analogy with Ref.~\cite{wang_experimental_2002}, we consider a driving scheme whereby one vortex is dragged by an external potential through the system, which is otherwise in equilibrium at a chosen temperature. Our results indicate that while the theorem seems largely unaffected by the sign of the temperature, the long-range interactions do appear to give rise to anomalously large entropy production over short timescales, leading to a disagreement with the FT prediction. However, over longer timescales the FT---and in turn, the second law of thermodynamics---is recovered. 

The rest of this paper is organised as follows. Section~\ref{sec:methods} describes the system setup and numerical methods employed. Section~\ref{sec:fluctuation_theorem} formally defines the Evans--Searles FT, and outlines our method for applying it to the point-vortex fluid. In Sec.~\ref{sec:results}, we discuss our main results. We explore how the FT predictions are affected by the temperature of the system, the finite system size, and noise added to the vortex dynamics. Finally, in Sec.~\ref{sec:conclusion} we conclude and discuss potential future research directions.

\section{Methods \label{sec:methods}}

\subsection{Point-vortex system}

We model a system of $\Nv$ point-vortices in a square geometry of side length $L$, with periodic boundaries in both directions. We fix $\Nv=100$ unless otherwise stated. Writing the position of vortex $i$ as $(x_i, y_i)$, and its circulation as $\Gamma_i = s_i \Gamma_0$ (with $s_i = \pm 1$, and $\Gamma_0$ a unit of circulation), the pseudo-Hamiltonian for such a system can be expressed~\cite{weiss_nonergodicity_1991}
\begin{equation}
H = -\sum_{i=1}^{\Nv}\sum_{j\neq i}^{\Nv}\frac{\Gamma_i \Gamma_j}{2}G(x_{ij},y_{ij}),
 \label{eq:sqarHamil}
\end{equation}
where
\begin{align}
 G(x_{ij},y_{ij}) = &\sum_{n = -\infty}^{\infty} \ln \left (\frac {\cosh[(2\pi x_{ij}/L) - 2 \pi n] - \cos(2\pi y_{ij}/L)}{\cosh(2\pi n)}\right) \nonumber \\  &-2\pi (x_{ij}/L)^2.
 \label{eq:h}
\end{align}
Here, $x_{ij} = x_i - x_j$ (and likewise for $y_{ij}$). We restrict our analysis to the neutral vortex system, for which there are an equal number of clockwise and anticlockwise circulating vortices. 

The conservative equations of motion for the vortices (which we denote with the superscript $^{(0)}$) are given by Hamilton's equations,
\begin{align}
    \Gamma_i \dot{x}_i^\mathrm{(0)} &= \partial H / \partial y_i \nonumber \\
    \Gamma_i \dot{y}_i^\mathrm{(0)} &= -\partial H / \partial x_i,
\end{align}
which for the Hamiltonian~\eqref{eq:sqarHamil} reduce to:
\begin{align}
    \dot{x}_i^\mathrm{(0)} &= \frac{\pi}{L}\sum_{j\neq i}^{\Nv} \left(\Gamma_j\sum_{n = -\infty}^{\infty}\frac{\sin(2\pi y_{ij}/L)}{\cosh[(2\pi x_{ij}/L)- 2 \pi n] - \cos(2\pi y_{ij}/L)}\right) \nonumber \\
    \dot{y}_i^\mathrm{(0)} &= -\frac{\pi}{L}\sum_{j\neq i}^{\Nv}\left( \Gamma_j\sum_{n = -\infty}^{\infty}\frac{\sin(2\pi x_{ij}/L)}{\cosh[(2\pi y_{ij}/L) - 2 \pi n] - \cos(2\pi x_{ij}/L)}\right).
\label{eq:eqmotion}
\end{align}

For the purposes of testing the fluctuation theorem in this system, we modify the dynamical model by adding a driving term in analogy with an earlier work~\cite{wang_experimental_2002}. Specifically, we attach one vortex to a harmonic trapping potential, which is translated at a constant velocity $\textbf{v}_\mathrm{trap} = v_\mathrm{trap} \hat{\bf e}_x$ across the system, where $\hat{\bf e}_x$ is the $x$-directional unit vector. 
In addition to the forces arising from interactions with other vortices, this `test' vortex thus experiences a restoring force $\mathbf{F}_\mathrm{restoring} = -K(\mathbf{r}_\mathrm{test} - \mathbf{r}_\mathrm{trap})$ towards the centre of the translating harmonic trap. Here, $\textbf{r}_\mathrm{test}$ ($\textbf{r}_\mathrm{trap}$) is the location of the test vortex (trap centre), and $K$ is a parameter that determines the stiffness of the trap. We assume that this force is realised by a dissipative Gaussian laser beam, which locally depletes the superfluid mass density from its bulk value $\rho$, pulling the test vortex towards its centre with a velocity $\dot{\mathbf{r}}_\mathrm{restoring} = \textbf{F}_\mathrm{restoring} / \rho \Gamma_0$~\cite{ao_berrys_1993, jackson_vortex_1999, groszek_motion_2018, simula_vortex_2018, foonote_force}. We account for this additional velocity by modifying the equations of motion~\eqref{eq:eqmotion} such that
\begin{align}
    \dot{x}_i &= \dot{x}_i^{(0)} - k (x_\mathrm{test} - x_\mathrm{trap}) \delta_{i,\mathrm{test}} \nonumber \\
    \dot{y}_i &= \dot{y}_i^{(0)} - k (y_\mathrm{test} - y_\mathrm{trap}) \delta_{i,\mathrm{test}},
\end{align}
where we have defined the spring constant $k = K/\rho \Gamma_0$. Numerically, we implement the restoring force in such a way that the dynamics are `unwrapped' with respect to the periodic boundaries: when the trap moves infinitesimally from $x=+L/2$ to $x=-L/2$, the force on the vortex does not change in strength or direction.

\begin{figure}[t]
\centering
\includegraphics[width=\columnwidth]{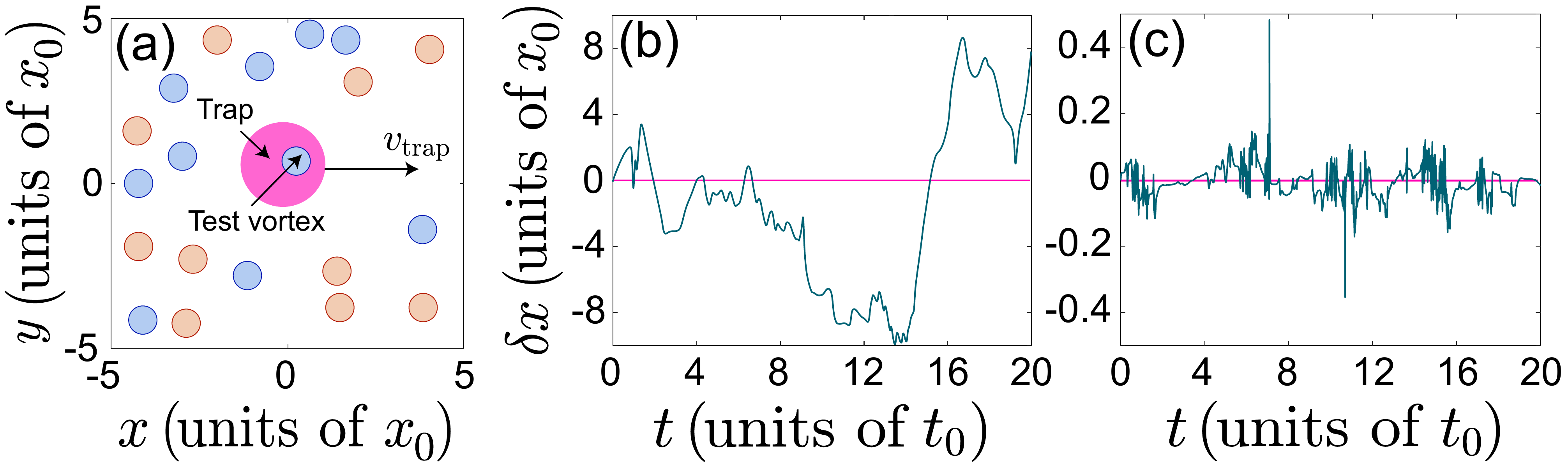}
\caption{Schematic of our system with $10$ positive circulation (blue) and $10$ negative circulation (red) point-vortices in a doubly periodic square geometry. The harmonic trapping potential (indicated by the pink shaded circle) moves at a constant speed $v_{\rm{trap}}$ in the positive $x$-direction, dragging the test vortex through the system. Panels (b) and (c) show the horizontal deflection $\delta x$ (blue line) of the test vortex relative to the trap centre (pink line) as functions of time for trap spring constants $k = 0.5\;t_0^{-1}$ and $k = 100\;t_0^{-1}$, respectively. In both cases, $v_{\rm{trap}} = 0.3$ $x_{\rm{0}}/t_{\rm{0}}$.}
\label{fig:system model FT}
\end{figure}

A schematic of our system setup is shown in Fig.~\ref{fig:system model FT}(a), with vortices (antivortices) shown as blue (red) circles, and the harmonic trap depicted by the pink shaded region with the test vortex captured inside of it. We track the trajectory of the test vortex during the dynamics, and use this to investigate the FT predictions for the vortex system (see Sec.~\ref{sec:fluctuation_theorem}). In Fig.~\ref{fig:system model FT}(b)--(c), we plot the horizontal displacement $\delta x$ of the test vortex (blue curve) relative to the trap center (pink line) as a function of simulation time, for two different spring constants $k$. As expected, the test vortex is able to drift farther from the trap center for smaller $k$ [panel (b)]. In all our simulations, the dynamics are initialised with the trap at the location of the test vortex (hence $\delta x = 0$ at $t = 0$), and the trap is translated at a velocity of $v_\mathrm{trap} = 0.3 x_0 / t_0$. Here and throughout this work we express length, time and energy in units of $x_0 = L/10$ and $t_0 = x_0^2 / \Gamma_0$, and $\varepsilon_0 = \rho \Gamma_0 x_0^2 / t_0$, respectively.

\subsection{Numerical implementation}

In the following, we explore the behaviour of the vortex system as a function of the (inverse) vortex temperature $\beta$~\cite{onsager1949statistical}. To achieve this, we sample the initial states for our dynamical simulations from a canonical ensemble at fixed $\beta$. 
As in earlier works~\cite{groszek_vortex_2018, valani_einsteinbose_2018, sharma_machine-learning_2022}, we achieve this using a Markov chain Monte Carlo method. Briefly, every step in the Markov chain involves randomly selecting one vortex from the configuration, attempting to move it a small distance in a random direction, and then deciding whether to accept the move. The probability of accepting a given move is given by the Metropolis rule, $\min \lbrace 1, \exp(-\beta \Delta H) \rbrace$, where $\Delta H$ is the change in the energy \eqref{eq:sqarHamil} that would be produced by the move. To avoid singular behaviour, we reject all moves that cause any two vortices to be separated by less than $0.0016 \; x_0$.

For a given choice of temperature, we perform a total of $10^6$ Markov chain steps. Following an initial burn-in of $10^5$ steps, we sample $1000$ microstates separated by intervals of $900$ steps to ensure minimal correlations between sampled states. 
We then use these microstates as our ensemble of initial conditions at the chosen $\beta$, and evolve each in time by numerically integrating Eq.~\eqref{eq:eqmotion} with the additional trapping potential described in the previous section. The dynamical simulations are conducted using a fourth-order Runge--Kutta method, with $6000$ numerical timesteps over an integration time of $40\; t_{\rm{0}}$.

\section{Fluctuation theorem \label{sec:fluctuation_theorem}}
The Evans--Searles fluctuation theorem predicts that, for a nonequilibrium finite system, the second law of thermodynamics will be violated over short timescales~\cite{evans_probability_1993, evans_equilibrium_1994}. Mathematically, the theorem states that over a time interval $\tau$, the probability $P(\sigma_\tau)$ of observing a phase-space trajectory that \textit{produces} entropy $\sigma_\tau$ is related to the probability $P(-\sigma_\tau)$ of observing a trajectory that \textit{consumes} an equivalent amount of entropy via the expression:
\begin{equation}
\frac{P(-\sigma_\tau )}{P(\sigma_\tau )} = \exp{(-\sigma_\tau)}.
\label{eq:FT}
\end{equation}
Since $\sigma_\tau$ is an extensive quantity, this ratio becomes increasingly small as either the system size or the time interval $\tau$ are increased, and hence the second law is recovered in the thermodynamic limit~\cite{wang_experimental_2002}.

Here we consider an integrated form of the FT~\cite{ayton_local_2001, wang_experimental_2002},
\begin{equation}
\frac{P(\sigma_\tau < 0)}{P(\sigma_\tau> 0)} = \langle
\exp{(-\sigma_\tau)}\rangle_{\sigma_\tau> 0},
\label{eq:FT2}
\end{equation}
where the angular brackets on the right-hand side (RHS) denote an average over all trajectories that produce entropy. The left-hand side (LHS) of Eq.~\eqref{eq:FT2} may be measured by taking the ratio of the number of entropy-consuming $(\sigma_\tau < 0)$ and entropy-generating $(\sigma_\tau > 0)$ trajectories over time interval $\tau$.

Our primary goal is to investigate the applicability of Eq.~\eqref{eq:FT2} for the case of point-vortices, by comparing the two sides of Eq.~\eqref{eq:FT2}. As a measure of the entropy production (or consumption) $\sigma_\tau$ produced over a time $\tau$ by the translating trap, we define the entropy production as a ratio of the work $W_\tau$ done by the translating trap to the thermal energy $k_\mathrm{B}T_\mathrm{p}$:
\begin{equation}
    \sigma_\tau = \frac{W_\tau}{k_\mathrm{B} T_\mathrm{p}}.
    \label{eq:sigma_general}
\end{equation}
Here, $T_\mathrm{p}$ is an ambient `phonon' temperature, which we treat as a free parameter in our simulations because the point-vortex model does not account for the dynamics of phonon degrees of freedom that would be present in a superfluid. We calculate the work $W_\tau$ done by the trap over time $\tau = t_f - t_i$ as an integral of the scalar product between the trapping force $\textbf{F}_\mathrm{restoring}$ acting on the test vortex and the trap translation velocity $\textbf{v}_\mathrm{trap}$. Hence Eq.~\eqref{eq:sigma_general} becomes:
\begin{equation}
\sigma_\tau = \alpha \int_{t_i}^{t_f} \diff s \mathbf{v}_\mathrm{trap}\cdot{\mathbf{F}_\mathrm{restoring}},
\label{eq:sigma}
\end{equation}
where we have defined the phonon (inverse) temperature $\alpha = 1/(k_{\rm{B}} T_{\rm{p}})$. Since the trap is translating at a constant velocity, this expression will result in entropy production ($\sigma_\tau > 0$) whenever the test vortex is behind the trap, and entropy consumption ($\sigma_\tau < 0$) when the test vortex gets pushed ahead of the trap due to interactions with other vortices in the system. 

\section{Results \label{sec:results}}

\subsection{Effect of the vortex temperature \label{sub:varying_vortextemp}}
We first consider a test of the FT as a function of the vortex temperature. To this end, we have run dynamical simulations for a range of initial temperatures spanning from the Berezinskii--Kosterlitz--Thouless (BKT) transition temperature $\beta_{\rm{BKT}} =  8\pi/\rho\Gamma_0^2$, at which the vortices and antivortices pair strongly to form dipoles~\cite{berezinskii_destruction_1971, berezinskii_destruction_1972, kosterlitz_ordering_1973}, to the Einstein--Bose condensation transition temperature $\beta_{\rm{EBC}} = -16\pi/\rho\Gamma_0^2 N_\mathrm{v}$, where the vortices arrange into same-sign clusters to maximise the energy~\cite{kraichnan_inertial_1967, kraichnan_two-dimensional_1980, viecelli_equilibrium_1995, valani_einsteinbose_2018}. In the following, we scale all positive temperatures $\beta_+$ by $\beta_\mathrm{BKT}$, and all negative temperatures $\beta_-$ by $\beta_\mathrm{EBC}$~\cite{groszek_vortex_2018, johnstone_evolution_2019, sharma_machine-learning_2022}.

\begin{figure}[!b]
\centering
\includegraphics[width=1\columnwidth]{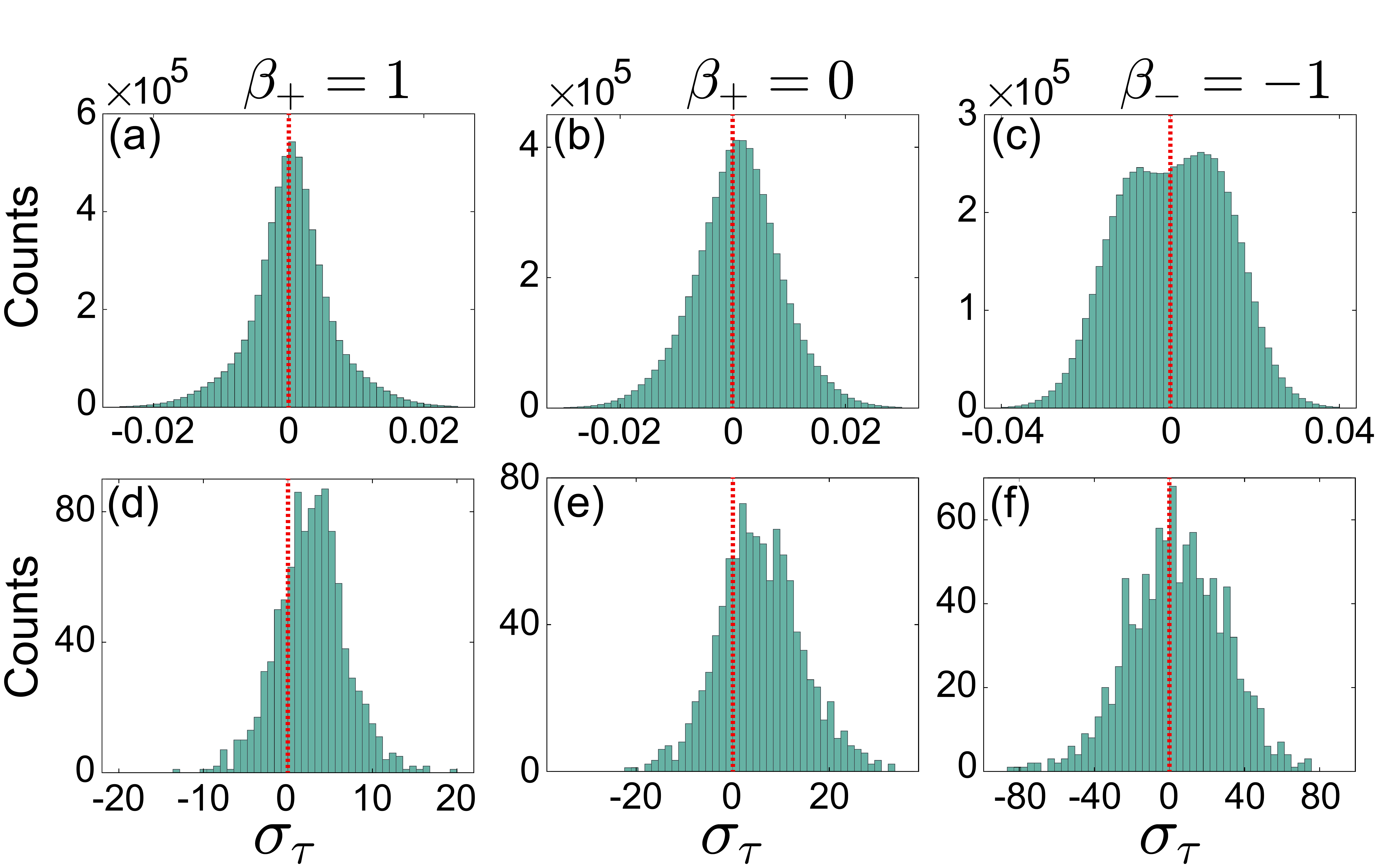}
\caption{Histograms of the dimensionless entropy production $\sigma_\tau$. The three columns correspond to inverse vortex temperature $\beta_+ = 1$, $\beta_+ = 0$ and $\beta_- = -1$, respectively. The measurement intervals are $\tau \approx 0.007\;t_0$ [(a)--(c)] and $\tau = 40\;t_0$ [(d)--(f)], and the dashed red vertical line in each panel indicates the location of $\sigma_\tau = 0$. Each histogram is produced from an ensemble of $1000$ computational trajectories, and we use every possible window of length $\tau$ in each simulation. The total sample sizes for these histograms are therefore $1000\times6000$ for (a)--(c), and $1000\times1$ for (d)--(f).}
\label{fig:hist FT}
\end{figure}

\begin{figure*}[t]
\centering
\includegraphics[width= 1.9\columnwidth]{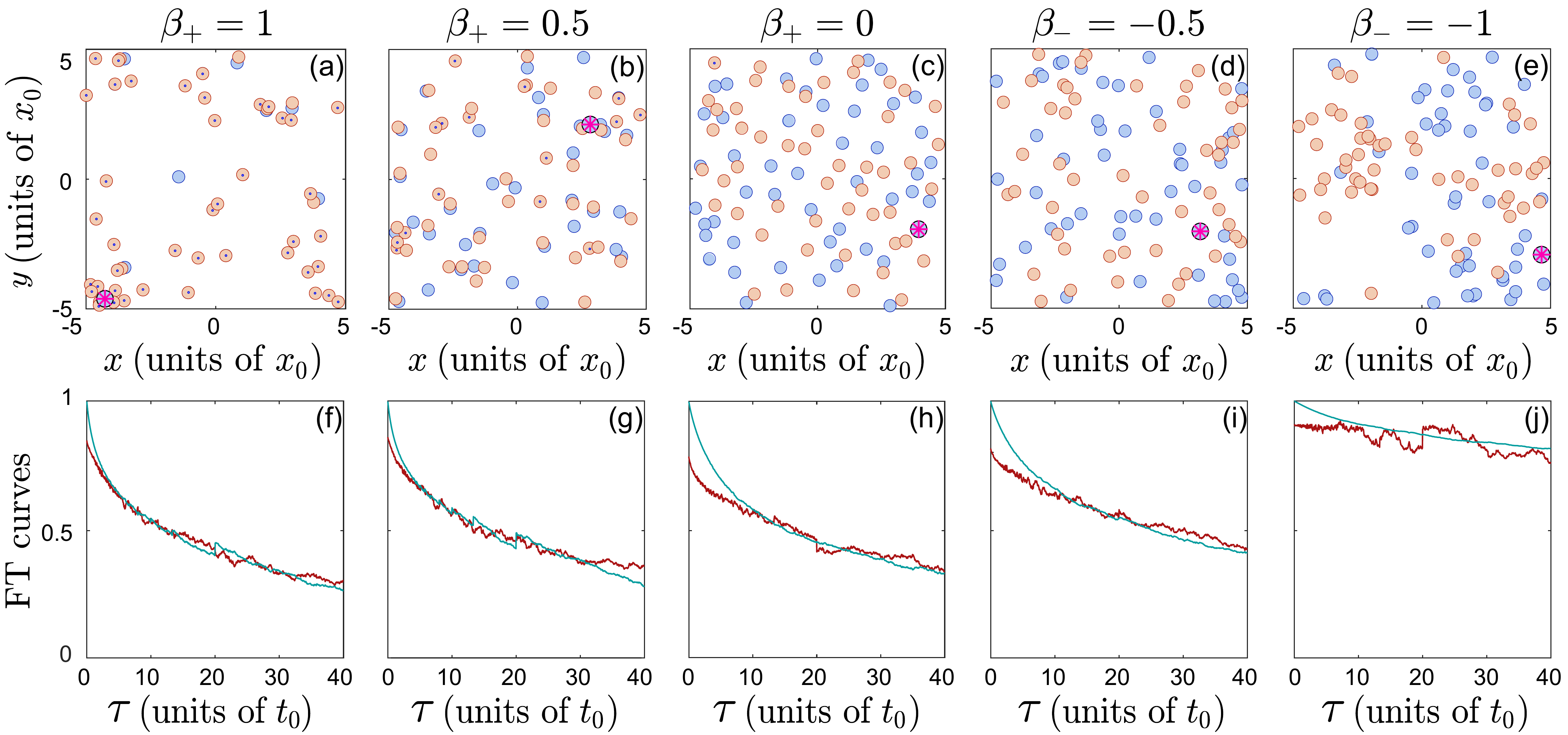}
\caption{Initial vortex configurations (a)--(e) and the corresponding fluctuation theorem (FT) curves (f)--(j) for inverse vortex temperatures $\beta_+ = \lbrace 1, 0.5, 0 \rbrace$ and $\beta_- = \lbrace -0.5, -1 \rbrace$, respectively. The blue and red markers in the top row represent the vortex and antivortex locations, respectively. A blue dot inside a red marker indicates the presence of a vortex directly beneath the antivortex. The pink asterisk indicates the trap location at $t = 0$, which coincides with the initial position of the test vortex. The bottom row shows the number ratio (red curves) of entropy consuming ($\sigma_\tau <0$) to entropy-producing ($\sigma_\tau> 0$) trajectories as per the LHS of Eq.~\eqref{eq:FT2}, together with the RHS of Eq.~\eqref{eq:FT2}, $\langle\exp{(-\sigma_\tau)}\rangle_{\sigma_\tau > 0}$ (teal curves). Each curve has been averaged over $1000$ simulations of a system with $N_\mathrm{v} = 100$. The trap parameters are $v_{\rm{trap}} = 0.3\;x_{\rm{0}}/t_{\rm{0}}$, $k = 100\;t_0^{-1}$.}
 \label{fig:test Ft}
\end{figure*}

In Fig.~\ref{fig:hist FT} we present histograms of the entropy production $\sigma_\tau$ at three vortex temperatures, $\beta_+ = 1$, $\beta_+ = 0$ and $\beta_- = -1$. For each temperature, we have produced histograms using both a short time interval, $\tau \approx 0.007$ [Fig.~\ref{fig:hist FT}(a)--(c)], and a longer interval $\tau = 40 \;t_{\rm{0}}$ [(d)--(f)]. The dashed vertical line in each panel denotes $\sigma_\tau = 0$. In panels (a)--(c), the distributions are almost symmetric about $\sigma_\tau = 0$, indicating that entropy consuming and producing trajectories are approximately equally likely for such short time intervals. By contrast, for the longer integration times shown in Fig.~\ref{fig:hist FT}(d)--(f), the histograms become skewed towards $\sigma_\tau > 0$, reflecting the tendency for entropy to be produced over long times, on average. Ultimately, for sufficiently long time intervals, entropy producing trajectories should become overwhelmingly dominant with almost vanishing probability of entropy-consuming trajectories, in accordance with the second law of thermodynamics. It can also be seen in Fig.~\ref{fig:hist FT} that as the vortex temperature shifts from positive to negative, the entropy distribution widens. This is presumably due to the stronger flow fields produced by the vortex clusters, which push the test vortex further from the trap centre, in turn giving rise to larger restoring forces. 

Using these entropy distributions, we can test the FT prediction in Eq.~\eqref{eq:FT2}. For the temperatures $\beta$ we have considered, we independently measure the LHS and RHS of Eq.~\eqref{eq:FT2} for varying time intervals in the range $0.007\;t_0 \lesssim \tau \leq 40\;t_0$. The results are presented in Fig.~\ref{fig:test Ft}. Panels (a)--(e) depict example initial vortex configurations at five inverse temperatures, $\beta_+ = \lbrace 1, 0.5, 0 \rbrace$ and $\beta_- = \lbrace -0.5, -1 \rbrace$, demonstrating the transition from dipole pairing to same-sign clustering as $\beta$ is reduced. The trap position is shown as a pink asterisk, which coincides with the test vortex at time $t=0$. Figure~\ref{fig:test Ft}(f)--(j) show the resulting FT curves corresponding to each temperature as a function of $\tau$, with the red (teal) line corresponding to the LHS (RHS) of Eq.~\eqref{eq:FT2}. Note that the right-hand side involves the free parameter $\alpha$, defined in Eq.~\eqref{eq:sigma}. We treat $\alpha$ as an optimisation parameter, and set it equal to the value for which the mean squared error between the two curves is minimised over all $\tau$. In all cases, it can be seen that the two curves start near unity and tend towards zero with increasing $\tau$, in broad agreement with the predictions of the fluctuation theorem. However, as $\beta$ is reduced, the timescale required for entropy production to dominate over entropy consumption increases. This suggests that at negative temperatures, our driving protocol becomes much less efficient at producing entropy, and instead continues to produce almost equal numbers of entropy-producing and entropy-reducing trajectories even for large $\tau$ [this is also reflected in the near-symmetry of the histogram in Fig.~\ref{fig:hist FT}(f)]. Regardless, Eq.~\eqref{eq:FT2} still appears to be broadly satisfied for $\beta < 0$, suggesting that the fluctuation theorem still holds even in this exotic temperature regime.

\begin{figure}[t]
\centering
\includegraphics[width= 1\columnwidth]{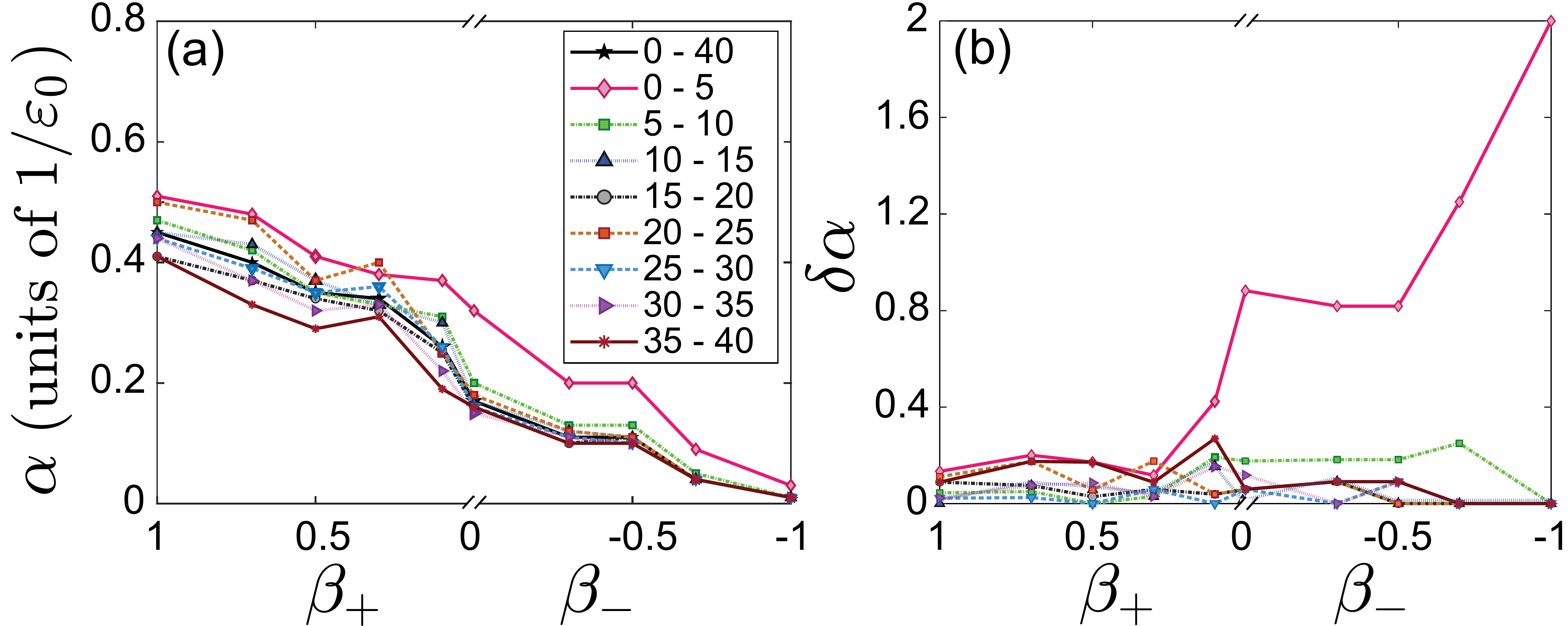}
\caption{(a) The fitted phonon inverse temperature $\alpha = 1/(k_\mathrm{B} T_\mathrm{p})$ as a function of the vortex temperature $\beta$. At each value of $\beta$, we have extracted $\alpha$ using nine different fitting time intervals, indicated in the legend. (b) The relative deviation $\delta \alpha = |\alpha - \alpha_0| / \alpha_0$ between each fitted $\alpha$ and the value $\alpha_0$ obtained from the fit to the full time window $\tau=[0$--$40]\;t_{\rm{0}}$. The color coding is as in (a).}
\label{fig:alpha beta plot}
\end{figure}

Curiously though, Fig.~\ref{fig:test Ft}(f)--(j) all show a slight disagreement between the two FT curves for small time intervals $\tau$. Specifically, the LHS of Eq.~\eqref{eq:FT2} (red curves) is lower than the RHS for small $\tau$, indicating that even for the shortest intervals $P(\sigma_\tau>0) > P(\sigma_\tau<0)$ in this system. Expressed another way, our point-vortex system never produces equal numbers of entropy-producing and entropy-reducing trajectories, even for arbitrarily small $\tau$. The value of $\tau$ at which the two curves first coincide increases weakly as $\beta$ is reduced, suggesting that this effect is at least partially dependent on the vortex temperature. We explore this discrepancy further in the following sections.

First, however, we investigate how the value of the fitted phonon temperature $\alpha$ varies as a function of the vortex temperature $\beta$. To ensure that its value is robust to the chosen window of $\tau$ over which we choose to fit the two sides of Eq.~\eqref{eq:FT2}, we measure $\alpha$ from fits to nine time intervals: $\tau \in \lbrace$0--40, 0--5, 5--10, 10--15, 15--20,  20--25, 25--30, 30--35, 35--40$\rbrace t_0$. The results are shown in Fig.~\ref{fig:alpha beta plot}(a). Interestingly, there is a near-linear relationship between $\alpha$ and $\beta$ (note that this trend continues across $\beta=0$ despite the difference in scaling for $\beta > 0$ and $\beta < 0$). However, $\alpha$ appears to be strictly positive, unlike $\beta$. Figure~\ref{fig:alpha beta plot}(b) shows the relative deviation $\delta \alpha = |\alpha - \alpha_0| / \alpha_0$ of each measured $\alpha$ from the value $\alpha_0$ extracted over the full fitting time interval $[0$--$40]\;t_{\rm{0}}$. Fitting to any interval beginning after $\tau \approx 10\;t_0$ gives $\alpha \approx \alpha_0$ (i.e.~near zero deviation). However, the strong deviation for the earliest time interval $[0$--$5]\;t_{\rm{0}}$ clearly quantifies the disagreement between the two FT curves for small $\tau$, which becomes more significant as $\beta$ is reduced towards increasingly negative temperatures.

\subsection{Finite-size effects}

The discrepancy between the two sides of Eq.~\eqref{eq:FT2} identified in Figs.~\ref{fig:test Ft} and \ref{fig:alpha beta plot} may be due to the finite size of our numerical simulation domain, in which case it should diminish as the system size increases and the thermodynamic limit is approached. To test this, we explore the effects of varying both the trap strength $k$ and the vortex number $\Nv$. Larger $k$ values prevent the test vortex from traversing large distances across the domain, effectively making the (periodic) boundaries appear further away. Larger $\Nv$ values, on the other hand, result in higher vortex densities, which essentially correspond to larger system sizes (except for an overall change in timescales, since the mean vortex velocity also increases).

\begin{figure}[!t]
 \centering
 \includegraphics[width= 1\columnwidth]{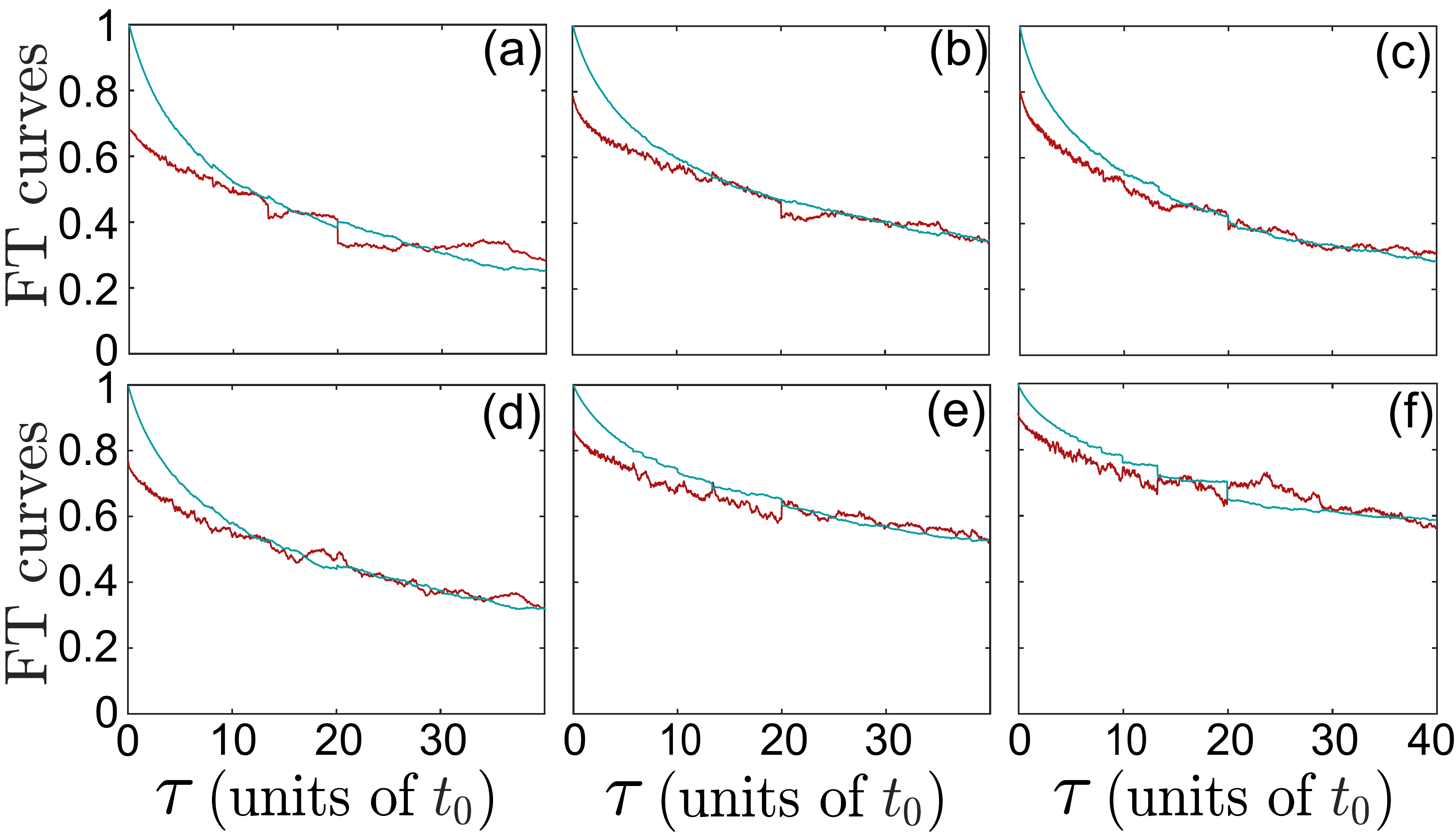}
 \caption{Tests of finite size effects at $\beta_+ = 0$. The top row shows the fluctuation theorem (FT) curves for a system with $N_\mathrm{v}=100$ vortices with trap strengths (a) $k = 7\;t_0^{-1}$, (b) $k = 100\;t_0^{-1}$, and (c) $k = 1000\;t_0^{-1}$. In the bottom row, the vortex number is (d) $\Nv = 50$, (e) $\Nv=200$ and (f) $\Nv=400$, with fixed trap strength $k = 100\;t_0^{-1}$. In each frame, the red (teal) curve corresponds to the left (right) side of Eq.~\eqref{eq:FT2}, as in Fig.~\ref{fig:test Ft}. In all cases, the trap speed is fixed at $v_{\rm{trap}} = 0.3$ $x_{\rm{0}}/t_{\rm{0}}$, and $\alpha$ values have been obtained by fitting to the full time interval (see Sec.~\ref{sub:varying_vortextemp}).}
\label{fig:finite size effect}
\end{figure}

Figure~\ref{fig:finite size effect} shows the results of our finite-size tests, with the vortex temperature fixed at $\beta_+ = 0$. The top row of Fig.~\ref{fig:finite size effect} shows the FT curves for $\Nv=100$ with trap strengths (a) $k = 7\;t_0^{-1}$, (b) $k = 100\;t_0^{-1}$, and (c) $k = 1000\;t_0^{-1}$. The deviation at small $\tau$ does appear to reduce as $k$ is increased, although the LHS of Eq.~\eqref{eq:FT2} (red curve) shows little indication of approaching unity at $\tau \approx 0$. It therefore does not appear that increasing $k$ is sufficient to completely eliminate the discrepancy. The bottom row of Fig.~\ref{fig:finite size effect} shows the FT curves for fixed trap strength $k=100\;t_0^{-1}$ and vortex numbers (d) $\Nv=50$ (e) $\Nv=200$, and (f) $\Nv=400$. The two curves appear to converge as $\Nv$ is increased, suggesting that the observed discrepancy may disappear as $\Nv$ is increased further. Nonetheless, it is interesting that this disagreement exists even in finite size systems, and hence we wish to explore its origin.

\subsection{Effect of long-range interactions}

Unlike earlier studies of the fluctuation theorem involving particles with contact interactions~\cite{wang_experimental_2002}, point-vortices are inherently long-range interacting. To investigate the importance of this feature of our system, here we introduce noise to the motion of the vortices, allowing us to effectively tune out the long-range interactions by overwhelming them with local fluctuations. Physically, this noise plays the role of the phonon bath in which the vortices would be immersed in a superfluid Bose--Einstein condensate. From the perspective of the test vortex, there are therefore two contributions to the environment it is moving through: a coherent part arising from long-range interactions, and an incoherent part corresponding to the noise. To explore the interplay between these two effects, we study three scenarios: (i) noise added to all vortices \textit{except} the test vortex, (ii) noise added to all vortices including the test vortex, and (iii) noise added to the test vortex when no other vortices are present. 
We implement the noise by adding an additional term, $\delta \textbf{v}_i = \eta_i \hat{\mathbf{e}}_x + \zeta_i \hat{\mathbf{e}}_y$, to the velocity $\mathbf{v}_i$ of vortex $i$ in Eq.~\eqref{eq:eqmotion}. The velocity increments $\eta_i$ and $\zeta_i$ are randomly generated each timestep from a uniform distribution within the range $[-\Delta, \Delta]$, where $\Delta$ is the chosen noise amplitude.

\begin{figure}[t]
   \centering
    \includegraphics[width= 1\columnwidth]{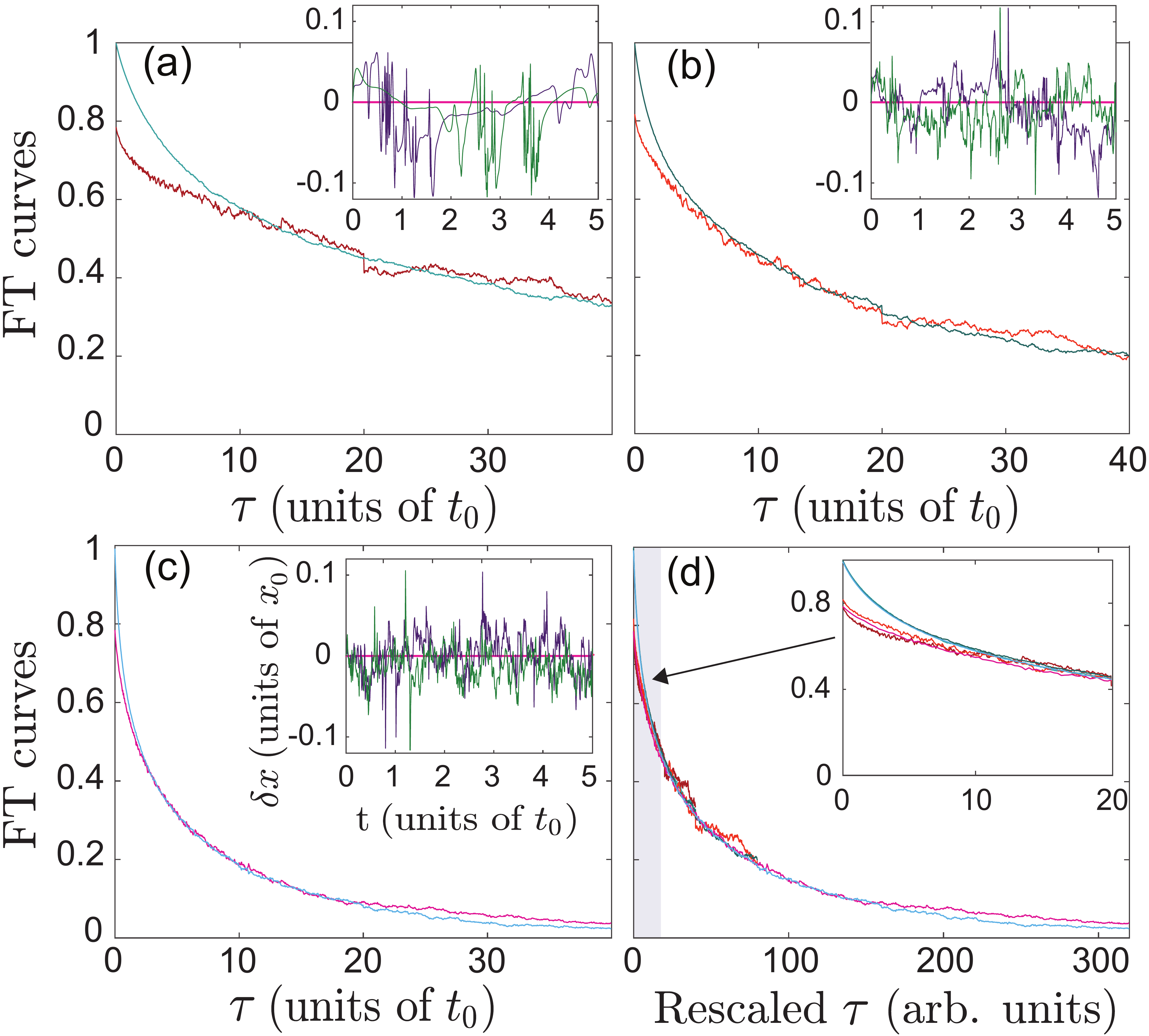}
    \caption{Fluctuation theorem (FT) curves at vortex temperature $\beta_+ = 0$ with noise added to the all vortices except the test vortex. The first three panels correspond to noise amplitudes (a) $\Delta=0$, (b) $\Delta = 100 \; x_0 / t_0$, and (c) $\Delta = 200 \; x_0 / t_0$. As in Fig.~\ref{fig:test Ft}, the red and teal curves correspond to the LHS and RHS of Eq.~\eqref{eq:FT2}, respectively. The insets of (a)--(c) each show two examples of the vortex deflection as in Fig.~\ref{fig:system model FT}(b)--(c). The axis labels for the insets in (a) and (b) are the same as for the inset in (c), but have been omitted for visual clarity. Frame (d) shows a collapse of the three datasets in (a)--(c), achieved by rescaling the time axis by multiplicative factors 1, 2 and 8, respectively. The inset of (d) shows a magnified view highlighting the small $\tau$ behaviour, with axes the same as for the main frame. In all cases, $v_{\rm{trap}} = 0.3\;x_{\rm{0}}/t_{\rm{0}}$, $k = 100\;t_0^{-1}$, and $\Nv = 100$.}
    \label{fig:noise effect plot}
\end{figure}

\begin{figure}[t]
    \centering
    \includegraphics[width= 1\columnwidth]{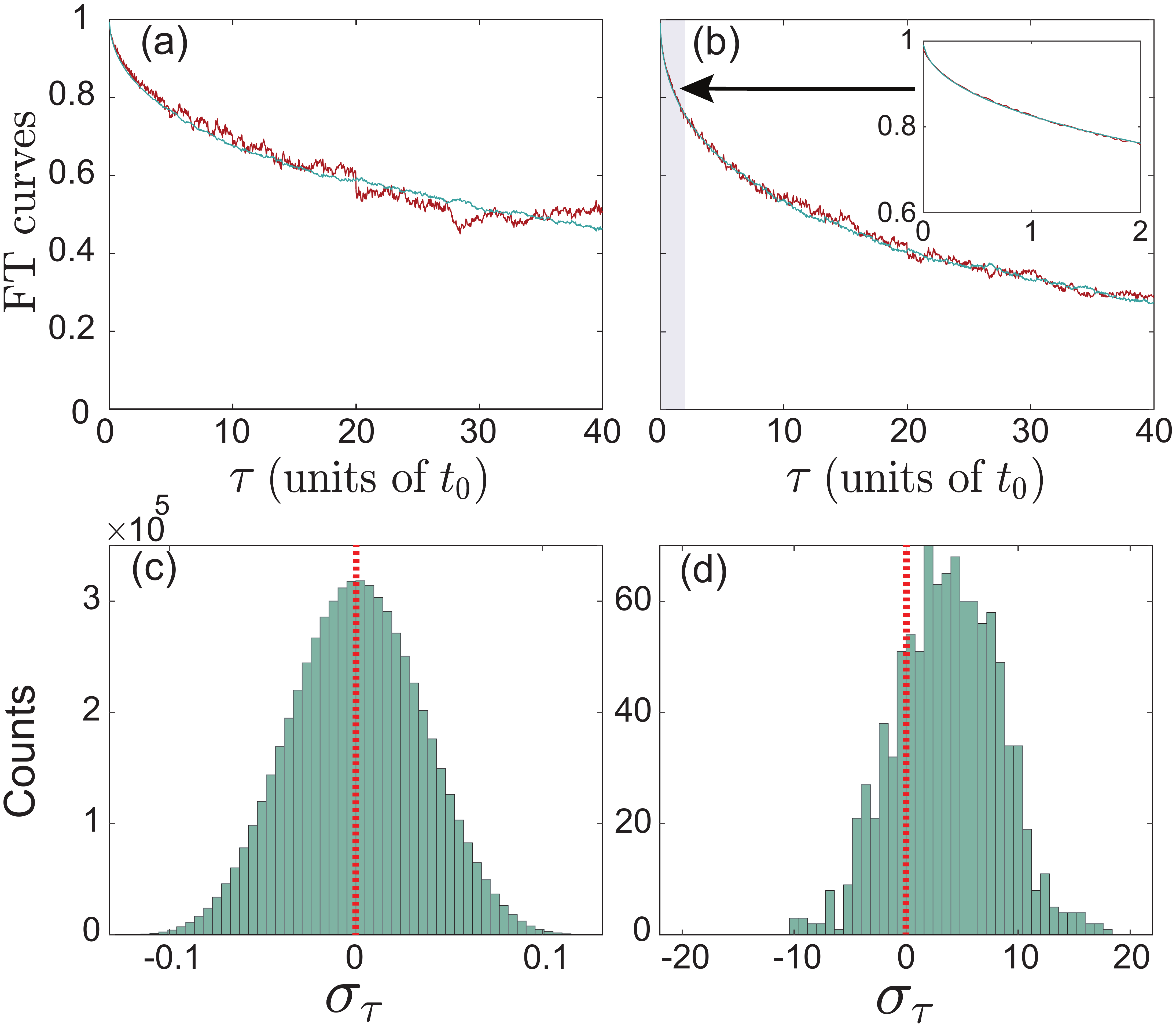}
    \caption{Fluctuation theorem results with noise added to the test vortex. (a) FT curves for a system of $\Nv = 100$ vortices at temperature $\beta_+ = 0$, with noise of amplitude $\Delta= 200 \; x_0 / t_0$ added to all vortices including the test vortex. Frames (b)--(d) correspond to a system with only the test vortex present, and added noise with amplitude $\Delta= 100 \; x_0 / t_0$. (b) FT curves, with inset showing an average over $10\,000$ simulations analysed for small $\tau$, as indicated by the purple shaded region. Panels (c) and (d) show histograms of the dimensionless entropy production $\sigma_{\tau}$ for time intervals $\tau = 0.007\;t_0$ ($1000\times6000$ samples), and $\tau= 40\;t_{\rm{0}}$ ($1000\times1$ samples), respectively. In all cases, $v_{\rm{trap}} = 0.3\;x_{\rm{0}}/t_{\rm{0}}$ and $k = 100\;t_0^{-1}$.
    }
    \label{fig:Single vortex FT}
\end{figure}

We first explore case (i), where noise is only added to the environment vortices. In Fig.~\ref{fig:noise effect plot}(a)--(c), we show the analysis of the two sides of Eq.~\eqref{eq:FT2} with noise amplitudes $\Delta = \lbrace 0, 100, 200 \rbrace x_0 / t_0$, respectively. Each panel includes an inset showing the deflection of the test vortex position from the trap center (horizontal pink line) as a function of time from two example simulations at the corresponding value of $\Delta$ (purple and green curves). At the outset it appears in Fig.~\ref{fig:noise effect plot}(b) and (c) that the early time deviation has been mitigated by the noise when compared with Fig.~\ref{fig:noise effect plot}(a). However, a careful analysis of panel (c) reveals that for very short time intervals the deviation persists. To make this observation clearer, Fig.~\ref{fig:noise effect plot}(d) reproduces the data in (a)--(c) with the time axis rescaled by factors of 1, 2, and 8, respectively. Under this rescaling, the data collapses, and hence increasing $\Delta$ in this scenario is effectively equivalent to reducing the timescale of the dynamics. In the inset of Fig.~\ref{fig:noise effect plot}(d), we focus on the small $\tau$ limit, clearly revealing that the deviation is present in all three cases. It therefore appears that no amount of noise added to the ``environment'' vortices could achieve agreement in this scenario. One possible explanation for this is that the long-range interactions are causing the deviation, meaning that the two curves would only coincide if local fluctuations were also added to the test vortex.

We next turn to case (ii), where noise is also added to the test vortex. This situation most closely resembles a true Bose--Einstein condensate, in which the phonon bath would affect all vortices equivalently. We have explored a range of noise amplitudes $\Delta$, and find that for $\Delta \lesssim 100 \; x_0 / t_0$, the deviation between the two sides of Eq.~\eqref{eq:FT2} at small $\tau$ persists. However, for noise amplitudes greater than this, the discrepancy is no longer visible. An example case with $\Delta = 200 \; x_0 / t_0$ is shown in Fig.~\ref{fig:Single vortex FT}(a). In this case, the left-hand side of Eq.~\eqref{eq:FT2} does approach unity as $\tau \to 0$, meaning that there equal numbers of entropy-producing and entropy-consuming trajectories in this limit. This supports the interpretation that the long-range interactions are responsible for the small $\tau$ anomaly, because at these amplitudes the noise is much stronger than the mean velocity $\bar{v}$ arising from long-range interactions, which is of order $\bar{v} \sim \Gamma_0/\bar{d} \sim 1 \; x_0 / t_0$ for our setup, where $\bar{d} \sim L / \Nv^{1/2}$ is the mean distance between vortices.

Finally, we examine case (iii), where only the test vortex is present and long-range interactions are entirely absent. This scenario trivially reduces to Brownian motion of the vortex in the trap, which more closely resembles earlier works on the fluctuation theorem~\cite{wang_experimental_2002}. The results of this test are presented in Fig.~\ref{fig:Single vortex FT}(b)--(d). Panel (b) shows the two sides of Eq.~\eqref{eq:FT2}, with an inset displaying data averaged over a larger ensemble. Evidently, the agreement is excellent for all $\tau$. This result can also be verified directly from the histogram in Fig.~\ref{fig:Single vortex FT}(c), which shows that the entropy production is distributed symmetrically around zero for the shortest time interval, $\tau=0.007\;t_0$, demonstrating an equal probability of positive and negative entropy trajectories. This in contrast to the $\tau = 40\;t_0$ case shown in Fig.~\ref{fig:Single vortex FT}(c), where the histogram is strongly skewed towards entropy production. Our results are therefore consistent with the explanation that the long-range interactions are responsible for the short time-interval deviations from the fluctuation theorem prediction of Eq.~\eqref{eq:FT2}.

\section{Conclusions \label{sec:conclusion}}

We have studied the fluctuation theorem in the context of a 2D vortex fluid by considering driven dynamics of an ensemble of point-vortices in a doubly periodic square domain at both positive and negative absolute vortex temperatures. We have found in general a good agreement with the predictions of the FT. However, for short time intervals, we have consistently observed anomalous deviations from the FT in our numerical simulations. These deviations were found to be persistent with respect to change of the finite system parameters, although they do appear to reduce as the vortex density was increased. Only when the long-range vortex--vortex interactions were either overwhelmed by noise or eliminated completely was a full agreement with the fluctuation theorem recovered. Hence we conclude that the long-range particle interactions in this system plausibly lead to anomalous deviations from the FT.

Our observations call for further investigations into Onsager's statistical hydrodynamics of point-vortices and into the role of long-range interactions in nonequilibrium systems more generally. In particular, it is known that nonequilibrium fluctuations in systems with short-range particle interactions readily generate long-ranged spatial correlations \cite{garrido_long-range_1990}. By contrast, our results point to a situation where long-range particle interactions appear to produce anomalous local entropy fluctuations. Given that the short time interval entropy production is found to exceed the FT prediction, we conjecture that this effect may potentially be explained by the trap indirectly dragging all vortices, mediated by the long-range interaction of the test vortex with the rest of the system vortices. Further elucidation of our observations may potentially have an impact on studies of quantum viscosity and non-equilibrium transport phenomena in superfluids.

\begin{acknowledgements}
This research was supported by the Australian Government through the Australian Research Council (ARC) Future Fellowship FT180100020, the ARC Centre of Excellence for Engineered Quantum Systems CE170100009, and the ARC Centre of Excellence in Future Low-Energy Electronics Technologies CE170100039.
\end{acknowledgements}


\begin{thebibliography}{48}
\bibitem [{foo()}]{foonote_force}%
  \BibitemOpen
  \href@noop {} {}\bibinfo {note} {A nondissipative potential would produce the
  same force, but the resulting velocity would be perpendicular to the
  potential, giving rise to circular motion around the trap
  centre.}\BibitemShut {Stop}%
\end{thebibliography}
\end{document}